\documentclass[11pt]{llncs}
\usepackage{graphicx, amsfonts, amsmath}
\usepackage{paralist}
\usepackage{xcolor}
\usepackage{pgfplots}
\usepackage{tikz}

\def\pms{\mathsf{pms}}
\def\msk{\mathsf{msk}}
\def\mpk{\mathsf{mpk}}
\def\pk{\mathsf{pk}}
\def\sk{\mathsf{sk}}
\def\Msg{\mathsf{Msg}}

\def\Sign{\mathsf{Sign}}
\def\Comb{\mathsf{Comb}}

\newtheorem{ibdtproto}{IBDT}
\newtheorem{proto}{Protocol}

\definecolor{bblue}{HTML}{4F81BD}
\definecolor{rred}{HTML}{C0504D}
\definecolor{ggreen}{HTML}{9BBB59}
\definecolor{ppurple}{HTML}{9F4C7C}

\title{Group Discounts Compatible with Buyer Privacy}
\author{Josep Domingo-Ferrer \and Alberto Blanco-Justicia}

\institute{Universitat Rovira i Virgili\\
Dept. of Computer Engineering and Mathematics\\
UNESCO Chair in Data Privacy\\
Av. Pa\"{\i}sos Catalans 26\\
E-43007 Tarragona, Catalonia\\
\email{\{josep.domingo,alberto.blanco\}@urv.cat}}

\begin{document}
\maketitle

\begin{abstract}
We show how group discounts can be offered without
forcing buyers to surrender their anonymity, as long
as buyers can use
their own computing devices ({\em e.g.} smartphone, tablet or computer) to
perform a purchase. Specifically, we present a protocol for privacy-preserving
group discounts. The protocol allows a group of buyers to prove
how many they are without disclosing their identities. Coupled with
an anonymous payment system, this makes group discounts
compatible with buyer privacy (that is, buyer anonymity).\\
\\
\textbf{Keywords:} Buyer privacy, Group discounts, Cryptographic protocols,
Digital signatures
\end{abstract}

\section{Introduction}\label{ab.sec.intro}
Group discounts are offered by vendors to encourage
consumers to use their services, to promote more efficient
use of resources, to protect the environment, etc. Examples include
group tickets
for museums, stadiums or leisure parks, discounted highway tolls
or parking fees for high-occupancy
vehicles, etc.
It is common for the vendor to require all group members to identify
themselves, but in reality this is seldom strictly necessary.

We make the assumption that the important feature about the group is the
{\em number of its members}, rather than their identities.
A secondary feature that may often (not always) be relevant for
a group discount is whether group members are physically together.

Anonymously proving the number of group members
and their being together is trivial in a face-to-face setting
with a human verifier,
who can see that the required number of people are present.
However, with an automatic verifier and/or in an on-line setting,
this becomes far from obvious.

In this paper, we propose a method
to prove the number of people in a group while preserving
the anonymity of group members and without requiring
specific dedicated hardware, except for a computing device
with some wireless communication capabilities ({\em e.g.}
NFC, Bluetooth or WiFi).
Also, we explore the option to include payment in our proposed
system, which is necessary for group discounts.
We complete the description of our method with a possible
anonymous payment mechanism, \emph{scratch cards}.
The method presented here is a generalization of
a specific protocol for toll discounts in high-occupancy
vehicles, whose patent we recently filed~\cite{hov}.

The rest of the paper is structured as follows.
Section \ref{ab.sec.buildingblocks} describes the building
blocks of our method, namely a digital
signature scheme, a key management scheme, an anonymous
payment scheme and wireless communication technologies;
the latter technologies should be short-range in applications
where one wants to check that the group members
are physically together.
Section \ref{ab.sec.system} describes our actual
group size accreditation method, including the required entities and protocols.
The security and the privacy
of our proposal are analyzed in Section~\ref{security}.
In Section \ref{ab.sec.implementation}, we give
a complexity estimation of our approach and describe
precomputation optimizations.
Finally, Section~\ref{ab.sec.conclusions} summarizes conclusions
and future work ideas.


\section{Building Blocks}\label{ab.sec.buildingblocks}

Our group size accreditation method is based on an identity-based
dynamic threshold (\emph{IBDT}) signature scheme, namely a
particular case of the second protocol proposed in~\cite{ab.Herranz2012}.

Threshold signature schemes are commonly based on
$(t,n)$-threshold secret sharing schemes, such as
the ones introduced in~\cite{ab.Blakley1979} and~\cite{ab.Shamir1979},
and they require a minimum number $t$ of participants
to produce a valid signature.
Dynamic threshold signature schemes differ from
the previous ones in that the threshold $t$ is not fixed during
the setup phase, but is declared at the moment of signing.
Our method takes advantage of this feature to find out
how many users participated in the signature of
a particular message, and consequently how many
people form a group.
If one wishes to prove that the signature is not only computed
by at least $t$ participants, but also that {\em these are
together in the same place}, the above signature schemes
need to be complemented
with short-range communication technologies.

On the other hand, identity-based public key signature schemes,
theorized by Shamir in~\cite{ab.Shamir1985} and with the
first concrete protocol, based on the Weil pairing,
developed by Boneh \emph{et al.} in~\cite{ab.Boneh2001},
allow public keys $\pk^U$
to be arbitrary strings of some length, which we call \emph{identities}.
These strings are associated with a user $U$ and reflect some
aspect of his identity, \emph{e.g.} his email address.
The corresponding secret key $\sk^U$ is then computed by a trusted entity,
the certification authority (CA), taking as input the user's identity and,
possibly, some secret information held only by the CA, and is
sent to the user $U$ through some secure channel.
Identity-based public key signature schemes offer a great flexibility
in key generation and management and our method takes advantage
of this feature by proposing a key management scheme that allows
preserving the anonymity of the participants.

Finally, in most group discounts, a fee must
be paid after proving the number of group members,
so an anonymous payment method is needed.
Indeed, this method should not reveal additional
information about the group members
to the service provider.


\subsection{IBDT Signature Scheme}\label{ab.subsec.ibdt}

We outline a general
identity-based dynamic threshold signature scheme, namely
the second protocol proposed in~\cite{ab.Herranz2012}.
Our protocol will be a slight modification of this general
case; we will point out differences when needed.
A general \emph{IBDT} signature scheme
consists of the following five algorithms.

\begin{ibdtproto}\label{ab.ibdtproto.setup}
\textbf{Setup} is a randomized trusted setup algorithm
that takes as input a security parameter $\lambda$,
a universe of identities $\mathcal{ID}$ and an integer $n$
which is a polynomial function of $\lambda$ and upper-bounds
the possible thresholds
(\emph{i.e.} $n$ is the maximum number of users that can participate
in a threshold signature).
It outputs a set of public parameters $\pms$
and a master key pair $\msk$ and $\mpk$.
An execution of this algorithm is denoted as
$$(\pms, \mpk, \msk) \leftarrow \mathsf{Setup}\left(\lambda, \mathcal{ID}, n\right).$$
\end{ibdtproto}

\begin{ibdtproto}\label{ab.ibdtproto.keygen}
\textbf{Keygen} is a key extraction algorithm that
takes as input the public parameters $\pms$,
the master key pair $\msk$ and $\mpk$,
and an identity $\mathbf{id} \in \mathcal{ID}$.
The output is a private key $SK_{\mathbf{id}}$.
An execution of this algorithm is denoted as
$$SK_{id} \leftarrow \mathsf{Keygen}\left(\pms, \mpk, \msk, \mathbf{id}\right).$$
\end{ibdtproto}

\begin{ibdtproto}\label{ab.ibdtproto.sign}
\textbf{Sign} is a randomized signing algorithm
that takes as input the public parameters $\pms$,
the master public key $\mpk$, a user's secret key $SK_{\mathbf{id}}$,
a message $\mathsf{Msg} \in \{0,1\}^{*}$ and
a threshold signing policy
$\Gamma = (t,S)$ where $S \subset \mathcal{ID}$ and $1 \le t \le |S| \le n$.
Note that, in our case, $t$ will be strictly equal to $|S|$.
Sign outputs a partial signature $\sigma_{\mathbf{id}}$.
We denote an execution of this algorithm as
$$\sigma_{\mathbf{id}} \leftarrow \mathsf{Sign}\left(\pms, \mpk, SK_{\mathbf{id}}, \Msg, \Gamma\right).$$
\end{ibdtproto}

\begin{ibdtproto}\label{ab.ibdtproto.comb}
\textbf{Comb} is a deterministic signing algorithm
which takes as input the public parameters $\pms$,
the master public key $\mpk$, 
the secret key of the
combiner user $SK_{\mathbf{id}}$, 
a message $\Msg$,
a threshold signing policy $\Gamma = (t,S)$ and
a specific set $S_t$ of $t$ partial signatures.
Comb outputs a global signature $\sigma$.
We denote the action taken by the signing algorithm as
$$\sigma \leftarrow \mathsf{Comb}\left(\pms, \mpk, SK_{\mathbf{id}}, \Msg, \Gamma, \{\sigma_{\mathbf{id}}\}_{\mathbf{id} \in S_t}\right).$$
\end{ibdtproto}

\begin{ibdtproto}\label{ab.ibdtproto.verify}
\textbf{Verify} is a deterministic verification algorithm
that takes as input the public parameters $\pms$,
a master public key $\mpk$, a message $\Msg$,
a global signature $\sigma$ and a threshold policy $\Gamma = (t,S)$.
It outputs $1$ if the signature is deemed valid and $0$ otherwise.
We denote an execution of this algorithm as
$$b \leftarrow \mathsf{Verify}\left(\pms, \mpk, \Msg, \sigma, \Gamma\right).$$
\end{ibdtproto}

For correctness, for any security parameter $\lambda \in \mathbb{N}$,
any upper bound $n$ on the group sizes, any universe $\mathcal{ID}$,
any set of public parameters and master key pair
$(\pms, \mpk, \msk)$,
and any threshold policy $\Gamma = (t,S)$ where $1 \le t \le |S|$, it is required that for
$$\sigma = \mathsf{Comb}\left(\pms, \mpk, SK_{\mathbf{id}}, \Msg, \Gamma, \{\sigma_{\mathbf{id}}\}_{\mathbf{id} \in S_t}\right),$$
$$ \mathsf{Verify}\left(\pms, \mpk, \Msg, \sigma, \Gamma \right) = 1$$
whenever the values $\pms$, $\mpk$, $\msk$
have been obtained by properly executing the
$\mathsf{Setup}$ algorithm, $|S_t| \ge t$, and for each
$\mathbf{id} \in S_t$,
$\sigma_{\mathbf{id}} \leftarrow \mathsf{Sign}(\pms, \mpk, SK_{\mathbf{id}}, \Msg, \Gamma)$ and
$SK_{\mathbf{id}} \leftarrow \mathsf{Keygen}(\pms, \mpk, \msk, \mathbf{id})$.


\subsection{Key Management}\label{ab.subsec.keymgnt}

The anonymity provided by our accreditation method is
a result of our key generation protocol and management solution.
As we stated above, identity-based public key cryptosystems
allow using arbitrary strings as public keys.
In our protocol, every user $U_i$ is given an ordered list
of public keys that
depend on some unique identifier of the user, such as his
national identity card number, his phone number, the
IMEI number of his phone or a combination of any of them.
We will call this identifier $n_{U_i} = d_{k}^{i} d_{k-1}^{i} \dots d_{1}^{i}$,
where $d_{j}^{i}$ is the $j$-th last digit of $n_{U_i}$
and typically ranges from $0$ to $9$.

To generate the list of public keys from an identifier $n_{U_i}$,
we choose a value $\ell < k$ and take the $\ell$ last digits
of $n_{U_i}$.
This results in a vector of public keys
$$\mathbf{PK}_{U_i}  = \left\{ \pk_{1}^{d_{1}^{i}} , \dots , \pk_{\ell}^{d_{\ell}^{i}}\right\},$$
with every $\pk_{j}^{d_{j}^{i}}$ being an encoding of the
digit and its position in $n_{U_i}$, for example:
$$\pk_{j}^{d_{j}^{i}} = j \, || \, d_{j}^{i},$$
where $||$ is the concatenation operation.
To illustrate this process, imagine $n_{U_i} = 12345678$
and $\ell = 4$. The resulting public key list would be
$$\mathbf{PK}_{U_i} = \left\{ 18, 27, 36, 45 \right\}.$$

To prove the number of members in a group,
the members will choose a common integer $j \in \{1,\dots , \ell\}$
so that the $j$-th public key in their list, \emph{i.e.}
$\pk_{j}^{d_{j}^{i}}$, is different for all of them.
Then they will perform the required operations with these public
keys and their corresponding private keys.
Assuming that the values of the digits range from
$0$ to $9$, this would provide anonymity to each
of the users, since on average 10\% of people
will share the same public key $\pk_{j}^{d_{j}^{i}}$
for some value of $j$.

Note that this approach limits the size of the groups
that can be certified with our method to a maximum
of 10. Moreover, intuition tells us that
the closer the size of the group to this maximum size,
the more difficult it becomes to find a value of $j$ for
which each user has a different public key.
The probability that our protocol fails depends
on the number of keys each user is given, $\ell$, and
the size of the group $n$; more specifically for
$n\le10$:
$$F(\ell,n) = \left( 1 - {10(10-1)\dots(10-n+1)\over10^n} \right)^\ell,$$
that is very close to $1$ for values of $n$ close to
$10$.

The limit on the maximum value of $n$ can be
increased by assigning $d\ge2$ digits of $n_{U_i}$ to each
of the $\ell$ public keys,
instead of just one digit.
By doing this, the maximum value for the size
of the groups becomes $10^d$, and the probability of
failure, for values of $n\le10^d$, is

$$F(\ell,n,d) = \left( 1 - {10^d(10^d-1)\dots(10^d-n+1)\over10^{dn}} \right)^\ell.$$
However, the price to be paid for choosing a larger
$d$ is a loss of anonymity,
since, if more digits are associated to each public key,
less users share the same public key. For example,
for $d=2$ a user would share each of his keys with
only $1\%$ of the total number of users.

The service provider will choose $\ell$ and $d$
depending on maximum number of keys that a user can store,
the maximum allowed group size and the anonymity level
to be guaranteed.


\subsection{Anonymous Payment Mechanisms}\label{ab.subsec.payment}

Group discounts are one of the applications of our method:
after proving the group size, the group
members must pay a fee that depends on that
size. If proving the size has been done anonymously,
it would be pointless to subsequently use a
non-anonymous payment protocol (such as credit card, PayPal, etc.).

Hence, we need to use an anonymous
payment mechanism along with our group size accreditation
protocol.
The simplest option for an anonymous payment method
is to use cash if the application and the service provider
allow it. Unfortunately, this will not always be the case,
and other payment methods have to be taken into account.
Electronic cash protocols such as~\cite{ab.Chaum1990}
are good candidates for this role.
Nowadays, Bitcoin~\cite{ab.Nakamoto2008}
is a well-established electronic currency and, although
it is not anonymous by design~\cite{ab.Reid2013},
it can be a good solution if accompanied by careful
key management policies.
Also, extensions of the original protocol as
Zerocoin~\cite{ab.Miers2013} provide
anonymity by design.

For completeness, we propose in this work to use a much 
simpler approach, based on prepaid scratch cards that users 
can buy at stores using cash (for maximum anonymity). 
Each such card contain a code $\mathsf{Pay.Code}$
which the service provider will associate with
a temporary account holding a fixed credit specified
by the card denomination. 


\subsection{Communication Technologies}\label{ab.subsec.comm}

Our accreditation method requires communication among
the members of a group and between the members and
some type of verifying device.
If we want to prove not only that a group has a
certain number of members, but also that these are
together, the interactions
with the verifying device must rely on
short-range communication technologies, like
NFC, Bluetooth or WiFi.

During the accreditation protocol, the users' smartphones
will be detected in some way by the verifying device and a
communication channel will be established. The requirements and
constraints of this process depend on the type of service
and verifying devices, but nonetheless it is desirable that
communication establishment be fast and not too cumbersome to the user.

We propose to use Bluetooth, and in particular
\emph{Bluetooth Low Energy}~\cite{ab.Bluetooth2013}
to communicate with the verifying device.
BLE solves some of the main limitations of traditional Bluetooth, \emph{i.e.}
reduces detection and  bonding times, requires much less work
by the user than NFC and has a shorter range than
both Bluetooth and WiFi, which is desirable in a method
like ours. Finally, BLE is implemented by most major
smartphone manufacturers, at least in recent models,
unlike NFC.

Regarding communication between the smartphones,
any of the three mentioned technologies, or a combination
of them (\emph{e.g.} Bluetooth pairing through NFC messages)
seems appropriate. The choice is up to
the service provider.


\section{Group Size Accreditation Method}\label{ab.sec.system}

A service that implements our accreditation method
includes the following elements:

\begin{itemize}
\item A service provider (SP) that publishes a smartphone
application $\mathsf{App}_U$ and distributes the
necessary public parameters and keys of an
\emph{IBDT} signature scheme $\Pi$ to users,
after some registration process.
\item A smartphone application $\mathsf{App}_U$ for each user $U$ which:
\begin{itemize}
\item allows computing signatures with $\Pi$ on behalf of $U$;
\item allows computing ciphertexts with a
public-key encryption scheme $\Pi'$ selected by
SP, under SP's public key $\pk^{SP}$;
\item can be run on master or slave mode,
which affects how $\mathsf{App}_U$
participates in the accreditation protocol.
\item includes some certificate which allows
checking the validity of $\pk^{SP}$;
\item implements some communication protocol, relying in
short-range communication technologies, such as NFC or
Bluetooth, to interact with the applications of the rest of
the members of the group and with the verifying devices.
\end{itemize}
\item Prepaid payment scratch cards available at stores.
Each card includes a code $\mathsf{Pay.Code}$ that the SP
associates to an account with a fixed credit specified by the card
denomination.
\item Verifying devices installed at suitable places in the provider's
infrastructures which:
\begin{itemize}
\item allow verifying signatures with $\Pi$;
\item hold the SP certificates as well as the keys needed to decrypt
ciphertexts produced with $\Pi'$ under $\pk^{SP}$.
\item have short-range communication capabilities and implement
some protocol to communicate with the users' devices.
\end{itemize}
\item Some method to penalize or prevent the misuse of the system.
\end{itemize}

The complete accreditation protocol runs as follows:

\begin{proto}
\textbf{System setup protocol.}
\begin{enumerate}
\item SP chooses the user identifier
to be used as $n_U$ and appropriate values for $\ell$ and $d$.
\item SP generates the parameters of the \emph{IBDT} signature
scheme $\Pi$ as per Algorithm $\mathsf{IBDT.Setup}$;
\item SP generates the parameters of
the public-key encryption scheme $\Pi'$.
\end{enumerate}
\end{proto}

\begin{proto}
\textbf{Registration protocol.}
\begin{enumerate}
\item A user $U$ with identifier $n_U$ authenticates
himself to the service provider, face-to-face or by some other
means. The user receives a PIN code $\mathsf{pin}_U$.
\item The service provider associates to $U$ a vector of public keys of $\Pi$,
$\mathbf{PK_{id}}$ as described in Section~\ref{ab.subsec.keymgnt}.
\item The service provider computes the secret keys
associated to $\mathbf{PK_{id}}$
as per Algorithm $\mathsf{IBDT.Keygen}$:
$$\mathbf{SK_{id}} = \left( sk_{1}^{d_1^\mathbf{id}}, \dots , sk_{\ell}^{d_\ell^\mathbf{id}} \right).$$
\item The user downloads the smartphone application $\mathsf{App}_U$ and, using
the PIN code $\mathsf{pin}_U$, completes the registration protocol and
receives the system parameters and keys, as well as the
public key $\mathsf{pk}^{SP}$.
\end{enumerate}
\end{proto}

\begin{proto}
\textbf{Credit purchase.}
\begin{enumerate}
\item A user buys a prepaid card for the system, \emph{e.g}. a scratch card,
from a store.
\item The card includes some code $\mathsf{Pay.Code}$ which has to be
introduced in the smartphone application.
\end{enumerate}
\end{proto}

\begin{proto}\label{ab.proto.gsetup}
\textbf{Group setup protocol.}
\begin{enumerate}
\item Some user $U^*$, among the group of users $U_1,\dots , U_t$ who want
to use the service, takes the leading role. This user
will be responsible for most of the
communication with the verifying device. $U^*$ sets his smartphone
application to run in master mode and the others set it to work in slave mode.
\item The users agree on a value $j \in \left\{ 1, 2, \dots , \ell \right\}$ such that the
value of the $j$-th public key in $\mathbf{PK_{id}}$ is different for every user.
\end{enumerate}
\end{proto}

\begin{proto}
\textbf{Group size accreditation protocol.}\label{ab.proto.accr}
\begin{enumerate}
\item A verifying device detects the users' devices and sends them
a unique time-stamped ticket $\mathsf{T}$ that may include
a description of the service conditions and options.
\item Each user $U_i$ runs Algorithm  $\mathsf{IBDT.Sign}$ to compute a
partial signature with $\Pi$ under his secret key $\sk_{j}^{d_j^i}$ on
message
$$\Msg = \left\langle \, \mathsf{T} \, || \, \pk_{j}^{d_j^1} \, || \, \dots \, || \, \pk_{j}^{d_j^t} \, \right\rangle,$$
for the threshold predicate $\Gamma = (t, \{ \pk_{j}^{d_j^1}, \dots , \pk_{j}^{d_j^t} \})$.
It sends the resulting partial signature $\sigma_\mathbf{i}$ to $U^{*}$.
\item $U^{*}$ receives $(\sigma_1 , \dots , \sigma_t )$ and runs
Algorithm  $\mathsf{IBDT.Comb}$
to combine these signatures and output a final signature $\sigma$ on behalf of $U_1 , \dots , U_t$.
$U^{*}$ sends to the verifying device
$$\mathsf{Msg'} = \left\langle \mathsf{Msg}, \sigma \right\rangle .$$
\item The verifying device checks the validity of the signature by running
$$\mathsf{IBDT.Verify} ( \Msg, \sigma , \pk_j^{d_j^1} || \dots || \pk_j^{d_j^t}, t) . $$
Note that this signature will only be valid if all users $U_1 , \dots , U_t$ have collaborated in
computing it, and thus it proves that the group of users is composed of at least $t$ people.
If the signature is not valid, the group will be penalized in
an application-dependent way, {\em e.g.} with access denial,
group discount denial, etc.
Otherwise, the service provider grants access to the
group of users and tells the group the amount $\mathsf{amount}_t$
they have to pay
depending on the group size.
\end{enumerate}
\end{proto}

\begin{proto}
\textbf{Payment.}
\begin{enumerate}
\item Each group member $U$ in the (sub)set $P$ of group members
who want to collaborate
in paying the bill sends to the verifying device via Bluetooth or WiFi his payment code encrypted
under SP's public key:
$$C_U = \mathsf{Enc}_{\mathsf{pk}^{SP}} ( \mathsf{T} || \mathsf{Pay.Code}_U ),$$
where $\mathsf{Pay.Code}_U$ is the code which user $U$ obtained from a prepaid scratch card
and where $\mathsf{Enc}$ is the public-key encryption algorithm
of scheme $\Pi'$.
\item The verifying device decrypts the
ciphertexts $\{ C_U : U \in P \}$ to obtain the payment
codes of the users in $P$.
\item The verifying device substracts the quantity $\mathsf{amount}_t$ divided by the cardinal
of $P$ to the accounts associated with the received payment codes.
\end{enumerate}
\end{proto}


\section{Security and Privacy Analysis}\label{security}

Security and privacy are offered by design in our proposal:
\begin{itemize}
\item The chosen IBDT scheme ensures unforgeability of
signatures under chosen message attacks even when an 
attacker can choose arbitrarily the threshold signing policy. 
In this case, this means that, for any $t \geq 2$,
no group of less than $t$ buyers is able to deceive the
service provider by producing a threshold signature with
threshold $t$. Complete security proofs can be found in
the original paper~\cite{ab.Herranz2012}.
\item No more than the pseudonyms and the number of participants of a group
is revealed to the service provider during the execution of the protocol. 
Buyer anonymity is guaranteed by the key management scheme
described in Section~\ref{ab.subsec.keymgnt} 
{\em within the community of buyers sharing the same public key}.
For example, if each public key is associated to a combination
of $d$ decimal digits, then on average this public key
is shared by a community containing $10^{-d} \times 100 \%$ of 
the total number of users.
\item When payment is completely anonymous, whatever anonymity
level achieved by key management is preserved after payment.
For our given method, this is ensured when a given $\mathsf{Pay.Code}$
cannot be linked to a specific buyer. This can be achieved, for example,
if the scratch card containing the $\mathsf{Pay.Code}$ is purchased using
cash.
\end{itemize}

\section{Performance Analysis}\label{ab.sec.implementation}
Our group size accreditation method is to be run by service
providers, specialized verifying devices and the
users' smartphones.
Therefore, it is important that the computations of the
underlying cryptographic protocol be as fast as
possible, especially the algorithms that are executed
by the smartphones, which have limited computational
capabilities and rely on batteries.

In this section, we analyze the performance of the
underlying \emph{IBDT} signature scheme.
This scheme is a
pairing-based cryptographic protocol and
as such, the required operations are performed in elliptic
curve groups.
We analyze its performance by counting the number
of point multiplications, point exponentiations and
pairings, which are the most costly operations.

Table~\ref{ab.tab.ibdttimes} shows the number of
these operations for each of the algorithms
in the \emph{IBDT} signature scheme.
The number of operations is counted as a function of the
maximum number of possible participants in a signature, $n$,
and the size of the signing group $t$.
As we stated previously, $t \le n$.

\begin{table}[!h]
\renewcommand{\arraystretch}{1.3}
\caption{Operations required per algorithm} \label{ab.tab.ibdttimes}
\centering
\begin{tabular}{c||c||c||c}
\hline
& \bfseries Multiplications & \bfseries Exponentiations & \bfseries Pairings \\
\hline
\hline
Setup & $0$ & $n+4$ & $1$\\
Keygen & $2n$ & $4n$ & $0$ \\
Sign	& $2n+6$	 & $2n + 5$ & $0$\\
Comb & $2n-t+1$ & $2n - t$ & $0$\\
Verify & $n+2$ & $n+1$ & $4$ \\
\hline
\end{tabular}
\end{table}

Note that the $\Sign$ and $\Comb$ algorithms, that are intended to be
executed in the users' smartphones during the
\textbf{group size accreditation protocol}~(\ref{ab.proto.accr}),
present what seems to be quite a high number of operations.
This might be a problem if the devices in which these algorithms
are to be executed do not have enough computational power.
Moreover, these two algorithms should precisely be
most efficient, since they are run most often, and possibly
with time constraints.
Therefore, it would be interesting if we could precompute
some of their operations.

The $\Sign$ algorithm is a probabilistic protocol, that is,
it has some random values in it that have to be
refreshed each time it is executed. This limits the amount
of operations in the algorithm that can be precomputed.
On the other hand, most of the operations depend on
static values, \emph{e.g.} keys and threshold policies $\Gamma$.
Threshold policies contain the number of signers that will
participate in a signature and their public keys.
We assume that groups of users will be quite stable,
\emph{i.e.} users will generally use services
together with the same
group members, or at least with a limited set of different groups.
We can exploit this assumption by precomputing operations
that only depend on static values and threshold policies.

The $\Comb$ algorithm obviously depends on the output
of $\Sign$, but it is a deterministic algorithm and some of
its operations depend on static values and also on the
threshold policies. Therefore, by the same assumption
as before, we can precompute some of the operations.

These precomputations will divide the $\Sign$ and $\Comb$
algorithms in two phases each, one for precomputing
values, which will be executed during the \textbf{group setup protocol}~(\ref{ab.proto.gsetup}),
and the other one performed during the \textbf{group size accreditation protocol}~(\ref{ab.proto.accr}).
The resulting number of operations in each of
these phases is shown in Table~\ref{ab.tab.ibdttimesprec}.

\begin{table}[!h]
\renewcommand{\arraystretch}{1.3}
\caption{Precomputed and non-precomputed
operations of the Sign and Comb algorithms (PC stands for
precomputed)} \label{ab.tab.ibdttimesprec}
\centering
\begin{tabular}{c||c||c||c}
\hline
& \bfseries Multiplications & \bfseries Exponentiations & \bfseries Pairings \\
\hline
\hline
Sign PC & $2n+2$ & $2n+1$ & $0$ \\
FastSign & $2$ & $4$ & $0$ \\
Comb PC & $2n-2t$ & $2n-2t$ & $0$ \\
FastComb & $3t+1$ & $3t$ & $0$ \\
\hline
\end{tabular}
\end{table}

\section{Conclusions and Future Work} \label{ab.sec.conclusions}
We have presented a privacy-preserving
mechanism for group discounts.
The method is built upon an \emph{IBDT} signature scheme,
a concrete key generation and management solution, short-range
communication technologies and anonymous payment mechanisms.
Our complexity analysis and initial tests show that the method 
is usable in practice.

Future work will consist of implementing
the protocol, testing it and 
developing a generic app for privacy-preserving group discounts that can be easily
customized for specific applications.


\section*{Acknowledgments}
The following funding sources are acknowledged:
Google (Faculty Research Award to the first author), 
Government of Catalonia (ICREA Acad\`emia Prize to the 
first author and grant 2014 SGR 537),
Spanish Government (project TIN2011-27076-C03-01 ``CO-PRIVACY''),
European Commission (FP7 projects
``DwB'' and ``Inter-Trust'') and Templeton World Charity
Foundation (grant TWCF0095/AB60 ``Co-Utility'').
The authors are with the UNESCO Chair in Data Privacy.
The views in this paper are the authors' own and 
do not necessarily reflect
the views of Google, UNESCO or the Templeton World Charity
Foundation.


\end{document}